# Noncentrosymmetric High-Temperature Superconductivity in doped $d^9$ Multiferroics

Hu Zhang

*Research Center for Computational Physics of Hebei Province, Hebei Key Laboratory of High-precision Computation and Application of Quantum Field Theory, College of Physics Science and Technology, Hebei University, Baoding 071002, People's Republic of China*

E-mails: zhanghu@hbu.edu.cn

**Abstract**

Multiferroics with $d^9$ electronic configurations, such as $SnCuO_2$, $PbCuO_2$, and $BiNiO_2$, exhibit coexisting antiferromagnetic order and ferroelectricity. Motivated by the fundamental link between symmetry breaking, strong electron correlations, and unconventional superconductivity, we propose a materials design strategy targeting noncentrosymmetric high-temperature superconductors through chemical doping of engineered $d^9$ multiferroics. This approach bridges two phenomena: (i) the coexistence of antiferromagnetism and ferroelectricity in correlated insulators, and (ii) the emergence of superconductivity in doped Mott/charge-transfer systems.

# I. Introduction

Superconductivity, as a quintessential macroscopic quantum coherent phenomenon, exhibits a profound intrinsic connection between its physical mechanism and symmetry. Conventional superconductors (such as those within the BCS theoretical framework) typically exist in crystal structures possessing spatial inversion symmetry, where the superconducting order parameter manifests as spin-singlet pairing. However, when the crystal structure lacks an inversion center (noncentrosymmetric crystal structure), the system exhibits unique superconducting behavior. Such materials are termed Noncentrosymmetric Superconductors (NCS). This broken symmetry induces strong antisymmetric splitting in momentum space via Spin-Orbit Coupling (SOC), significantly altering the electronic band topology. This key feature drives the superconducting pairing mechanism beyond the conventional paradigm: on one hand, strong SOC induces spin splitting of the bands (e.g., Rashba-type or Dresselhaus-type), destroying spin degeneracy; on the other hand, the superconducting order parameter may exhibit a novel state of mixed singlet-triplet pairing.

Prototypical systems, such as the heavy-fermion compound $CePt_3Si$, the layered superconductor $Li_2Pd_xPt_{3-x}B$, and the family of superconducting topological insulators (e.g., $PbTaSe_2$), have become frontier platforms for exploring novel quantum states of matter. Research on these materials holds significant theoretical importance for understanding the emergence mechanisms of superconducting pairing under symmetry constraints, and also demonstrates potential application value in fields such as topological quantum computation, spintronics, and low-dissipation devices.

High-temperature superconductivity in cuprates and nickelates emerges from doping their antiferromagnetic parent compounds, where strong electron correlations dominate. The parent phases of copper oxides (e.g., $La_2CuO_4$, $CaCuO_2$) and infinite-layer nickelates (e.g., $NdNiO_2$) feature $d^9$ configurations ($Cu^{2+}$: $3d^9$, $Ni^+$: $3d^9$), forming charge-transfer insulators with antiferromagnetic (AFM) order.

Previously, we have proposed the concept of $d^9$ multiferroics ($SnCuO_2$, $PbCuO_2$,

BiNiO$_2$) derived from high-temperature superconducting parent compounds [1,2]. Using density functional theory (DFT) and hybrid functional calculations, we demonstrated their coexistence of AFM order and ferroelectricity. d$^9$ multiferroics are good platforms to study physics of doping properties of electronic systems with a d$^9$ configuration. If superconductivity appears in doped d$^9$ multiferroics, noncentrosymmetric high-temperature superconductors can be obtained. We argue this question in this work.

## II. Results and Discussion

### A. The concept of d$^9$ multiferroics

Ferroelectric materials undergo a structural phase transition from a nonpolar, high-symmetry paraelectric state to a polar state. In this polar phase, the spontaneous electric polarization is switchable under an applied electric field. This behavior is fundamentally characterized by a double-well potential energy landscape, as schematically illustrated in Fig. 1(a).

On the other hand, the parent compounds of copper-based high-temperature superconductors—such as La$_2$CuO$_4$ and CaCuO$_2$—feature Cu$^{2+}$ ions with a d$^9$ electronic configuration. This results in nine electrons occupying the copper $d$ orbitals, leading to a half-filled Cu 3d$_{x^2-y^2}$ band, a pivotal electronic structure depicted in Fig. 1(b). Consequently, the CuO$_2$ planes in these undoped parent compounds exhibit a Mott insulating ground state with antiferromagnetic (AFM) order localized on the Cu sites. Doping these parent materials subsequently unlocks rich phenomena, most notably high-temperature superconductivity.

Our previous work aims to integrate this distinctive d$^9$ electronic structure into ferroelectric systems, thereby engineering a novel class of multiferroic materials. We anticipate that such hybrid systems will exhibit emergent physical properties arising from the interplay between ferroelectricity and strong electron correlations. This approach conceptually bridges the fields of ferroelectrics and strongly correlated d$^9$ systems. These materials are called d$^9$ multiferroics. In our previous work, we have predicted d$^9$ multiferroics including SnCuO$_2$, PbCuO$_2$ and BiNiO$_2$ derived from high-temperature superconducting parent compounds.

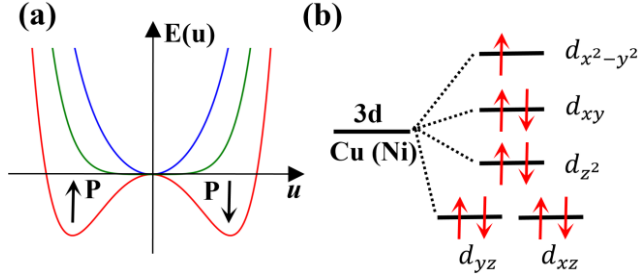

Fig. 1. Schematic diagrams of (a) ferroelectrics and (b) a $d^9$ configuration on Cu and Ni.

## B. BiNiO$_2$: A Mott multiferroic

In 2019, superconductivity was experimentally discovered in Sr-doped NdNiO$_2$, synthesized by reducing the perovskite NdNiO$_3$ to its infinite-layer phase NdNiO$_2$ with space group $P4/mmm$ (No. 123) [3]. The parent compound NdNiO$_3$ adopts an orthorhombic structure ($Pbnm$, No. 62), with experimental lattice parameters $a$ = 5.39 Å, $b$ = 5.38 Å, $c$ = 7.61 Å. Similarly, BiNiO$_3$ exhibits an orthorhombic phase ($Pbnm$) with comparable lattice parameters. Additionally, BiNiO$_3$ stabilizes in a triclinic phase (P-1, No. 2) with lattice parameters a = 5.39 Å, b = 5.65 Å, c = 7.71 Å and angles α = 92.182°, β = 89.780°, γ = 91.704°. Given the structural analogy between NdNiO$_3$ and BiNiO$_3$, we propose that BiNiO$_2$—the infinite-layer counterpart of BiNiO$_3$—should be experimentally accessible via analogous reduction of perovskite BiNiO$_3$.

We have theoretically predicted a Mott multiferroic phase in BiNiO$_2$. BiNiO$_2$ undergoes a ferroelectric structural transition from a nonpolar to a polar phase (space group $P4mm$, see Fig. 2b), exhibiting broken inversion symmetry, driven by the stereochemically active lone pair on Bi$^{3+}$ [1]. The lowest-energy ground state of polar BiNiO$_2$ is a Hubbard Mott insulator exhibiting G-type antiferromagnetic ordering. This substitution of Nd$^{3+}$ by Bi$^{3+}$ establishes a materials design pathway bridging parent compounds of nickelate superconductors and strongly correlated multiferroics.

The Bi$^{3+}$ 6s$^2$ lone pair induces a displacement of NiO$_2$ layers along the $c$-axis,

reducing the Ni-O bond length from 1.92 Å (nonpolar) to 1.87 Å (polar). HSE06 calculations show that nonmagnetic and ferromagnetic states remain metallic, with Bi-6*p* states contributing significantly to the Fermi surface. However, the G-type AFM state opens a Mott gap of 0.5 eV, splitting the Ni-$3d_{x^2-y^2}$ band into lower and upper Hubbard bands. The calculated electric polarization (0.49 C/m²) arises primarily from Bi displacements, enabling electric-field control of the Ni magnetic moments (0.93 μB). Thus, $BiNiO_2$ is a $d^9$ multiferroic.

**C. $SnCuO_2$ and $PbCuO_2$: Lone-pair-driven ferroelectricity**

In prior work, we also proposed a strategy for designing multiferroic materials with $d^9$ electronic configurations by leveraging magnetic ordering in copper-oxygen layers — a hallmark of copper oxide high-temperature superconductors — to induce ferroelectricity. Theoretical predictions identify tin copper oxide ($SnCuO_2$) and lead copper oxide ($PbCuO_2$) as candidate systems [2]. These charge-transfer multiferroics crystallize in the *polar P*4*mm* space group, exhibiting broken inversion symmetry. Crucially, the stereochemically active lone pairs of Sn and Pb hybridize with O 2p orbitals, driving buckling in $Cu-O_2$ layers and activating the lone-pair mechanism for ferroelectricity. With their $d^9$ configuration, $SnCuO_2$ and $PbCuO_2$ behave as charge-transfer insulators featuring antiferromagnetic ground states localized on Cu sites. This electronic structure preserves key strongly correlated properties characteristic of *parent compounds* in copper oxide superconductors. Our findings demonstrate a viable approach to engineering multiferroicity within high-temperature superconducting copper oxide frameworks.

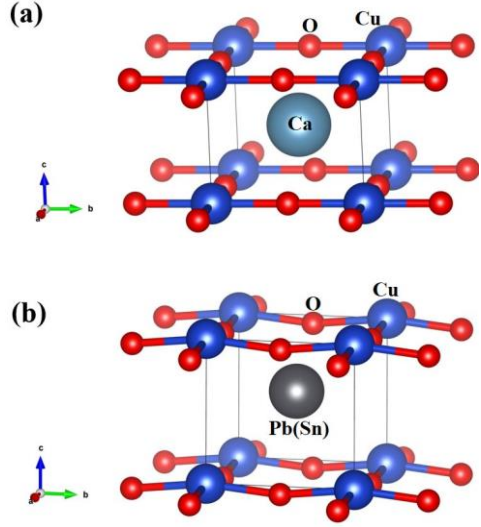

Fig. 2. Crystal structures of (a) CaCuO$_2$ with *P*4/*mmm* symmetry and (b) SnCuO$_2$ (PbCuO$_2$ and BiNiO$_2$) with *P*4*mm* symmetry.

The crystal structure of CaCuO$_2$ is shown in Fig. 2a. The infinite-layer structure of CaCuO$_2$ (*P*4/*mmm* symmetry) serves as the parent compound for designing SnCuO$_2$ and PbCuO$_2$. Replacing Ca$^{2+}$ with Sn$^{2+}$ or Pb$^{2+}$ introduces stereochemically active lone pairs (Sn-5s$^2$, Pb-6s$^2$), which hybridize with O-2p orbitals to induce out-of-plane buckling in CuO$_2$ layers (Fig. 2b). This buckling structure breaks inversion symmetry, stabilizing the polar *P4mm* phase. Structural optimization using PBE functional yields lattice parameters of $a$ = 3.78 Å, $c$ = 3.98 Å for SnCuO$_2$ and $a$ = 3.82 Å, $c$ = 3.73 Å for PbCuO$_2$. Phonon spectra of the nonpolar *P*4/*mmm* phase reveal unstable A$_{2u}$ polar modes (imaginary frequencies: −215 cm$^{-1}$ for SnCuO$_2$, −186 cm$^{-1}$ for PbCuO$_2$), confirming the soft-mode-driven ferroelectric transition. The double-well energy barrier for polarization switching is 0.4 eV in SnCuO$_2$ and 0.03 eV in PbCuO$_2$, suggesting lower coercivity for the latter. Figure 3 shows electronic structures of PbCuO$_2$ with *P4mm* symmetry. The Cu $3d_{x^2-y^2}$ band crosses the Fermi level that is the same as CaCuO$_2$. Along the Γ-Z direction, the Cu $3d_{x^2-y^2}$ band has weak dispersions showing a two-dimensional feature. Hence, PbCuO$_2$ still has a d$^9$ configuration.

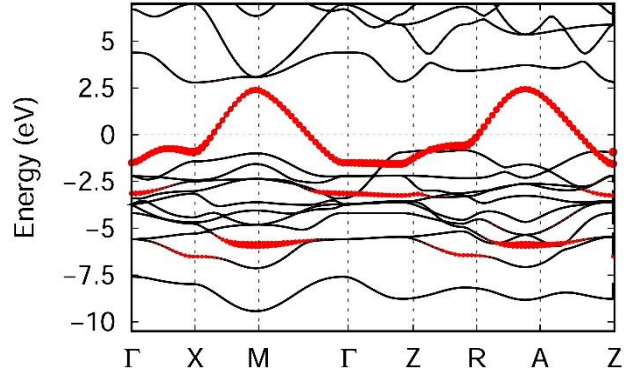

Fig. 3. Electronic structures of PbCuO$_2$ with *P4mm* space group with broken inversion symmetry.

Electronic structures calculated via HSE06 hybrid functional confirm the charge-transfer insulating nature of both materials. Magnetic calculations reveal G-type AFM order as the ground state, with intralayer exchange coupling $J_1$ = 76 meV (SnCuO$_2$) and 82 meV (PbCuO$_2$), comparable to undoped cuprates. In SnCuO$_2$, the Cu-3d$_{x^2-y^2}$ band lies 0.37 eV above the O-2$p$ valence band maximum (VBM), while PbCuO$_2$ exhibits a narrower gap of 0.28 eV. The partial density of states (PDOS) highlights strong hybridization between Sn/Pb-$s$/$p$ and O-2$p$ orbitals, critical for stabilizing the polar phase. SnCuO$_2$ and PbCuO$_2$ are copper-based d$^9$ multiferroics.

**D. Doped d$^9$ multiferroics**

The discovery of high-temperature superconductivity in doped copper oxides revolutionized condensed matter physics by revealing unconventional superconducting states beyond the conventional BCS paradigm. Central to this phenomenon is the role of chemical doping—a controlled introduction of charge carriers into parent compounds that are typically antiferromagnetic (AFM) Mott insulators as shown in Fig. 4. Mott Insulator Basis: Parent compounds (e.g., La$_2$CuO$_4$, Nd$_2$CuO$_4$, Bi$_2$Sr$_2$CaCu$_2$O$_8$) exhibit a d$^9$ electronic configuration with a half-filled Cu 3d$_{x^2-y^2}$ band, resulting in strong electron correlations and AFM order. Doping (hole/electron injection) disrupts long-range AFM, creating a metallic state that hosts superconductivity. A mysterious partial gap in electronic spectra above Tc in underdoped regimes, unrelated to superconductivity. Strange metal behavior: Linear-in-temperature resistivity (ρ∝T) in the normal state, violating Fermi-liquid theory. Competing orders: Charge/spin density waves, intra-unit-cell loop currents, and electronic nematicity.

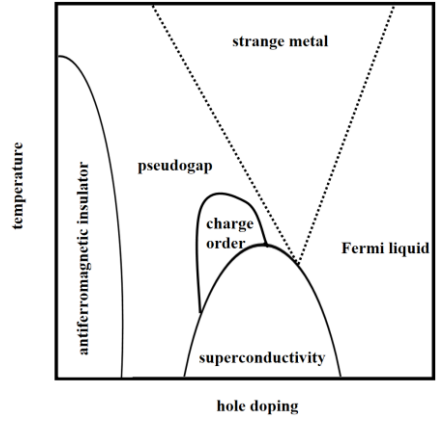

Fig. 4. Schematic diagram of doping regulation of high-temperature superconducting materials.

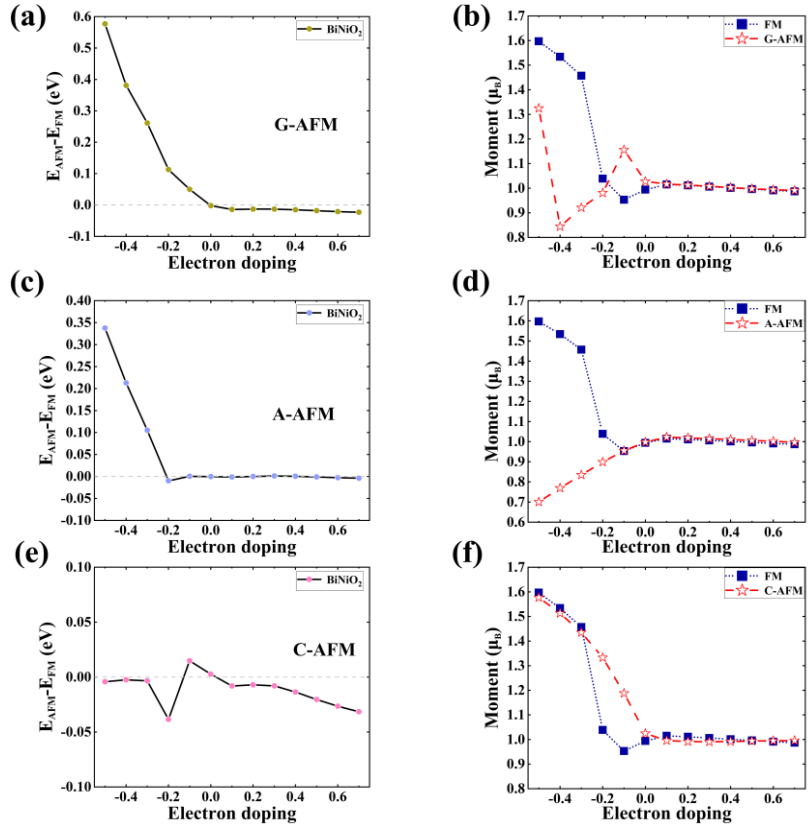

Fig. 5. DFT+U calculations of (a, b) the total energy difference and magnetic moments between G-type antiferromagnetic (G-AFM) and ferromagnetic (FM) orders, (c, d) A-type antiferromagnetic (A-AFM) and FM orders, and (e, f) C-type antiferromagnetic (C-AFM) and FM orders in $BiNiO_2$ at different doping concentrations.

Figure 5 and 6 show doping properties of BiNiO$_2$ and PbCuO$_2$. We observe that in BiNiO$_2$ with the G-AFM state, the energy difference between antiferromagnetic (AFM) and ferromagnetic (FM) states progressively approaches zero with increasing electron doping concentration. This indicates suppression of magnetic ordering. Under FM-state conditions: With electron doping, the magnetic moment stabilizes near 1 μ$_B$. With hole doping, the moment reaches a minimum at 0.1 e/f.u. hole concentration, then increases monotonically with further hole doping. In the AFM state: During electron doping, the moment remains stable near 1 μ$_B$. Under hole doping, the moment exhibits non-monotonic evolution: it first increases, then decreases, and finally increases again with rising hole concentration.

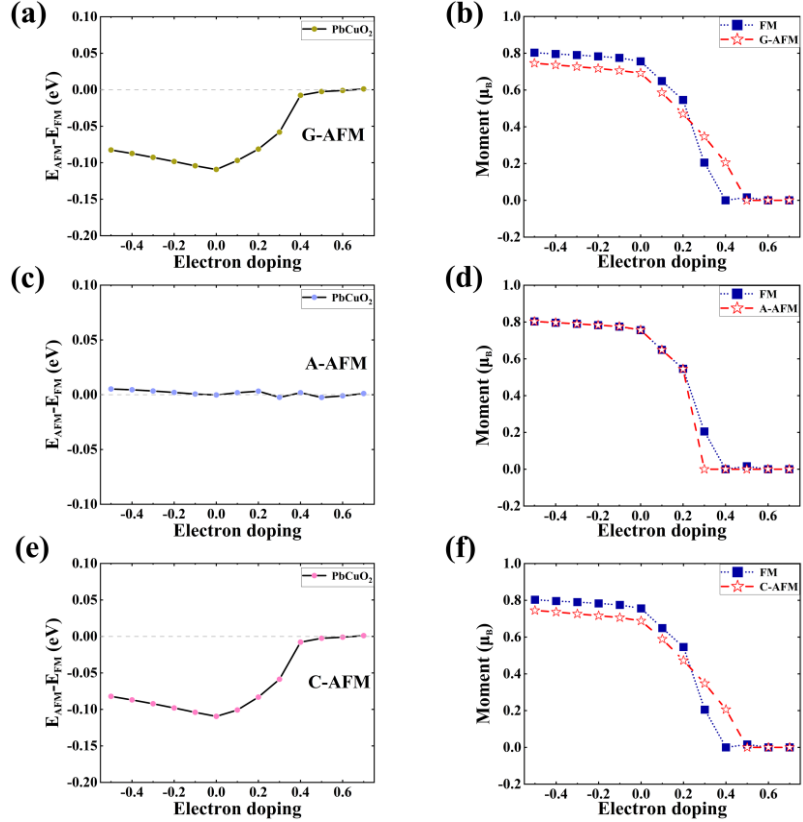

Fig. 6. DFT+U calculations of (a, b) the total energy difference and magnetic moments between G-type antiferromagnetic (G-AFM) and ferromagnetic (FM) orders, (c, d) A-type antiferromagnetic (A-AFM) and FM orders, and (e, f) C-type antiferromagnetic (C-AFM) and FM orders in PbCuO$_2$ at different doping concentrations.

**E. Seeking superconductivity in doped d⁹ multiferroics**

Recent advances reveal an untapped design space at the intersection of multiferroicity and high-temperature superconductivity. The core hypothesis posits that *d⁹ multiferroics*—exemplified by $SnCuO_2$, $PbCuO_2$, and $BiNiO_2$—offer a unique platform for realizing noncentrosymmetric superconductors. These materials intrinsically combine: Broken inversion symmetry (*P4mm* polar space group); Strong spin-orbit coupling (SOC) (heavy cations: $Sn^{2+}$/$Pb^{2+}$/$Bi^{3+}$); Mott/charge-transfer insulating ground states with robust antiferromagnetic (AFM) correlations. Chemical doping of these systems could simultaneously induce metallicity and preserve the critical symmetry conditions for unconventional pairing. Doped $d^9$ multiferroics represent a materials-by-design frontier where ferroelectric distortion, strong SOC, and Mott physics conspire to host high-Tc superconductivity. The recent synthesis of superconducting $NdNiO_2$ validates the core strategy of reducing perovskite nickelates. With $BiNiO_2$ thin films now within experimental reach, this platform offers the most viable path toward **high-temperature noncentrosymmetric superconductors**.

### III. Conclusions

In summary, we have discussed the basic physical properties of doped $d^9$ multiferroics. Due to the coexistence of inversion symmetry breaking and a $d^9$ electronic configuration, it is reliable to seeking superconductivity in doped $d^9$ multiferroics, which allow us to obtain noncentrosymmetric high-temperature superconductors. This provides an opportunity to study mixed singlet-triplet pairing in copper oxides and nickelate oxides.